\documentclass[a4paper,twocolumn,
english,aps,pre,floatfix,superscriptaddress,showpacs]{revtex4-1}
\usepackage[T1]{fontenc}
\usepackage[latin1]{inputenc}
\usepackage{amsmath}
\usepackage{babel}
\usepackage{graphics}
\usepackage{amssymb}


\begin{document}
\title
{Universal and non-universal amplitude ratios for scaling corrections
on Ising strips
}

\author {S. L. A. \surname{de Queiroz}}
\email{sldq@if.ufrj.br}
\affiliation{Rudolf Peierls Centre for Theoretical Physics, University of
Oxford, 1 Keble Road, Oxford OX1 3NP, United Kingdom}
\affiliation{Instituto de F\'\i sica, Universidade Federal do
Rio de Janeiro, Caixa Postal 68528, 21941-972
Rio de Janeiro RJ, Brazil}

\date{\today}

\begin{abstract}
We consider strips of Ising spins at criticality. For strips of  width $N$ sites,   
subdominant (additive) finite-size corrections to scaling are 
assumed to be of the  
form $a_k/N^k$ for the free energy, and $b_k/N^k$ for inverse correlation length,
with integer values of $k$. 
We investigate the set $\{a_k,b_k\}$ ($k \geq 2$) by  exact evaluation
and numerical transfer-matrix diagonalization techniques, and their changes
upon varying anisotropy of couplings, spin quantum number $S$, and (finite)
interaction range, in all cases for both periodic (PBC) and free (FBC) 
boundary conditions across the strip. We find that the coefficient
ratios $b_k/a_k$ remain constant upon varying coupling anisotropy for $S=1/2$
and first-neighbor couplings, for both PBC and FBC 
(albeit at distinct values in either case). 
Such 
apparently universal 
behavior is not maintained upon changes in $S$ or interaction range.  
\end{abstract}
\maketitle
 
\section{Introduction} 
\label{intro}
In this paper we investigate corrections to scaling in critical Ising systems
on a strip geometry. Consider a square
lattice with $N$ lines and $M$ columns, in the limit $M \to  \infty$.
Other two-dimensional lattices, such as triangular or honeycomb, can be
brought into a square-like shape, by suitable bond additions or deletions.
From the largest ($\Lambda_0$) and second-largest ($\Lambda_1$) eigenvalues
of the column-to-column transfer-matrix (TM), one obtains the free energy
per spin, $f_N$ (in units of $k_BT$), and spin-spin correlation length $\xi_N$, 
via~\cite{domb60}:
\begin{equation}
N\,f_N =\zeta\,\ln \Lambda_0 \ ;\qquad \xi_N^{-1}=\zeta\,\ln \frac{\Lambda_0}
{|\Lambda_1|}\ .
\label{eq:fNxiN}
\end{equation}
The factor $\zeta$ is unity for the square lattice and, in triangular and honeycomb
geometries (also for the square lattice when the TM progresses along the 
diagonal~\cite{baxter,obpw96}), 
corrects for the fact that the physical length added upon each application
of the TM differs from one lattice spacing~\cite{priv84}.
In all cases of interest here, i.e., ferromagnetic systems, $\Lambda_0$ 
and $\Lambda _1$ are both real and positive.

At the critical point $T_c$ where a second-order transition takes place, 
conformal invariance~\cite{cardy} gives the following relations 
regarding universal quantities $c$, the conformal anomaly~\cite{bcn86},
and the spin scaling dimension $x_1$~\cite{cardy84}:
\begin{eqnarray}
\lim_{N \to \infty}N^2\,(f_N-f_\infty) -2N\,f_{\rm surf}=\alpha\pi c \ ;
\label{eq:c}\\
\lim_{N \to \infty} N\,\xi_N^{-1}=\beta\pi x_1\ .\qquad\qquad
\label{eq:eta}
\end{eqnarray}
In Eq.~(\ref{eq:c}), where $c=1/2$ for models in the Ising 
universality class, $f_{\rm surf}=0$ for strips with periodic boundary conditions 
(PBC) across, and non-zero for free (FBC) or fixed BCs; $\alpha=\frac{1}{6}$
for PBC, and $\frac{1}{24}$ for FBC~\cite{bcn86}. 
In Eq.~(\ref{eq:eta}), where the exponent $x_1$ for the Ising universality class is
$x_1^b=\frac{1}{8}$  in the bulk, and $x_1^s=\frac{1}{2}$ for the ordinary surface 
transition, one has $x_1=x_1^b$, $\beta=2$ for PBC, 
and  $x_1=x_1^s$, $\beta=1$ for FBC~\cite{cardy84}.       

Since Eqs.~(\ref{eq:c}) and~(\ref{eq:eta}) are expected to be exact only
asymptotically, it is of interest to develop a systematic understanding
of the corresponding finite-$N$ corrections. We write:
\begin{eqnarray}
N\,(f_N-f_\infty) -2\,f_{\rm surf} =\sum_{k=1}^\infty \frac{a_k}{N^k}\ ,
\label{eq:c1}\\
\xi_N^{-1} =\sum_{k=1}^\infty \frac{b_k}{N^k}\ ,\qquad\qquad
\label{eq:eta1}
\end{eqnarray}
where $a_1=\alpha\pi c$, $b_1=\beta \pi x_1$.
Assuming only integer powers of $N^{-1}$ in Eqs.~(\ref{eq:c1}) and~(\ref{eq:eta1})
is believed to be warranted as long as one is dealing with models in the 
Ising universality class~\cite{cardy86}. 
We revisit this assumption in Sections~\ref{sec:s=1} and~~\ref{sec:nnn}
below. 
Our task here will be to learn as much as possible about
the coefficients $\{a_k,\,b_k\}$, $k \geq 2$, as well as 
(for reasons explained below) their ratios $b_k/a_k$. We are interested
in their respective universality, or lack thereof, upon changes in boundary conditions,
degree of spatial anisotropy of interactions, spin quantum number $S$, and (finite)
interaction range. We restrict ourselves to the square lattice.

In Section~\ref{sec:pbc} we investigate $S=1/2$ strips with PBC, first-neighbor
interactions, and varying degrees of spatial anisotropy;
in Sec.~\ref{sec:fbc}, we examine systems with FBC, again with varying
anisotropy; Sec.~\ref{sec:s=1} deals with the spin--$1$ case, and isotropic
couplings only; in Sec.~\ref{sec:nnn} we return to $S=1/2$ and introduce
next-nearest-neighbor couplings (keeping to isotropic interactions).
Finally, in Sec.~\ref{sec:conc}, concluding remarks are made.

\section{Periodic boundary conditions}
\label{sec:pbc}

\subsection{Preliminaries; isotropic systems}
\label{prelim}

We recall results for strips cut along the $x$ direction, 
with $N$ lines and $M \to \infty$ columns, and PBC across. All eigenvalues of the TM
can be written in closed form~\cite{domb60}. With $K_i \equiv J_i/k_BT$ being
the interactions respectively along $x$ ($i=1$) and $y$ ($i=2$), $\Lambda_0$ 
and $\Lambda_1$ are:
\begin{eqnarray}
\ln \Lambda_0 -\frac{1}{2}N \ln (2 \sinh 2K_1)=\frac{1}{2} \sum_{r=0}^{N-1} 
\gamma_{2r+1}\ ,\qquad\quad 
\label{eq:ev0}\\
\ln \Lambda_1 -\frac{1}{2}N \ln (2 \sinh 2K_1)=\frac{1}{2} \sum_{r=0}^{N-1} 
\gamma_{2r}\ , \qquad\qquad
\label{eq:ev1}
\end{eqnarray}
where
\begin{equation}
\cosh \gamma_r =\cosh 2K_1^\ast\,\cosh 2K_2-\sinh 2K_1^\ast\,\sinh 2K_2\,\cos 
\omega_r\ ;
\label{eq:gamma_r}
\end{equation}
the dual couplings $K_i^\ast$ are defined by $\tanh K_i^\ast=\exp (-2K_i)$,
and the allowed frequencies are $\omega_r=r\pi/N$.

With $s_i \equiv \sinh 2K_i$, one has  $s_1\,s_2=1$ at the critical temperature 
where the system is self-dual, and Eq.~(\ref{eq:gamma_r}) becomes:
\begin{equation} 
\cosh \gamma_r =1+\frac{1}{s_1^2}(1-\cos \omega_r)\qquad (T=T_c)\ .
\label{eq:gamma_r2}
\end{equation}
For isotropic systems, $s_1=s_2=1$ at criticality. In this case,
the sums in Eqs.~(\ref{eq:ev0}) and~(\ref{eq:ev1}) 
were tackled~\cite{izhu01}
by applying the extended Euler-MacLaurin summation formula~\cite{abram,ivizhu02}:
\begin{eqnarray}
\sum_{n=0}^{N-1}F(a+nh+\alpha h)=\frac{1}{h}\int_a^b F(x)\,dx +\qquad\qquad\qquad
\nonumber \\
+\sum_{k=1}^\infty \frac{h^{k-1}}{k!}\,
B_k(\alpha)\,\left(F^{(k-1)}(b)-F^{(k-1)}(a)\right)\ ,\qquad
\label{eq:e-ml}
\end{eqnarray}
where $h=(b-a)/N$, $F^{(j)}(x)$ is the $j$--th derivative of $F(x)$, 
$0 < \alpha <1$, and the $B_k(\alpha)$ are the periodic Bernoulli 
polynomials (related to the Bernoullli {\em numbers}, denoted simply by $B_k$,
by $B_k=B_k(0)$). 

It was found that only odd powers of $N^{-1}$, i.e., $k=2j-1$, $j\geq 1$ occur in
Eqs.~(\ref{eq:c1}) and~(\ref{eq:eta1}); this can be traced back to 
the fact that the Bernoulli numbers $B_m$ obey 
$B_{2m-1}=0$ $(m>1)$. Also, relatively 
simple closed-form expressions were derived for all $a_k$ and $b_k$. 
Such expressions reproduce previously-known exact results 
(for $a_1$~\cite{bcn86}, $b_1$~\cite{cardy84}, and $b_3$~\cite{dds82}),
and are in very good agreement with numerically-obtained ones~\cite{dq00}.
Furthermore, although the coefficients themselves are non-universal
(upon changing lattice structure, or considering quantum Ising 
chains~\cite{pfeuty,fs78,hb81} instead
of their two-dimensional classical counterparts), their ratio
is found to remain 
constant upon the same set of changes~\cite{izhu01}:
\begin{equation}
\frac{b_k}{a_k}=\frac{2^{k+1}-1}{2^k-1}\quad (k=2j-1\ ,\ j\geq 1)\ [\,{\rm PBC}\,].
\label{eq:akbk}
\end{equation}

It should be noted that when one considers the TM running along the diagonal 
of the square lattice (as in Refs.~\onlinecite{baxter,obpw96}), 
one gets for isotropic systems with PBC the same value for the ratio $b_k/a_k$ as in
Eq.~(\ref{eq:akbk}). Furthermore, the coefficients themselves have
the same absolute value as those found with the TM along $x$; only, they
alternate in sign: $a_k$, $b_k <0$ for $k=1$, positive for $k=2$ etc~\cite{izpriv2}.

\subsection{Anisotropic systems with PBC}
\label{anispbc}

With $K_1/K_2 \equiv R \neq 1$, one gets~\cite{inw86,nb83} the corresponding forms of 
Eqs.~(\ref{eq:c}) and~(\ref{eq:eta}) [$\,$specializing to Ising spins on strips with 
PBC$\,$] as:
\begin{eqnarray}
\lim_{N \to \infty}N^2\,(f_{N(i)}-f_\infty)=\frac{1}{s_i}\,\frac{\pi}{12}\ ,
\label{eq:c3}\\
\lim_{N \to \infty} N\,[\,\xi_{N(1)}^{-1}\,\xi_{N(2)}^{-1}\,]^{1/2}=\frac{\pi}{4}\ ,
\label{eq:sqrt}
\end{eqnarray}
where $f_{N(i)}$  and $\xi_{N(i)}$ are, respectively, free energy and 
correlation length at criticality, both calculated by iterating the
TM along the direction with couplings $K_i$.
Note~\cite{inw86} that $f_\infty$ in Eq.~(\ref{eq:c3}) also
depends on $R$~. $s_1 \equiv \sinh 2K_c$ is the solution of
$\sinh 2K_c\,\sinh 2RK_c=1$.

As noted in Ref.~\onlinecite{izhu01}, Eq.~(\ref{eq:gamma_r2}) can be rewritten as:
\begin{equation}
\gamma(\omega)=2\,\ln (u+\sqrt{1+u^2})\ ,\qquad 
u \equiv \frac{1}{s_1}\sin \frac{\omega}{2}\ .
\label{eq:gamma_r3}
\end{equation}
In this  form, it is immediate to see that anisotropy brings about a simple
rescaling of the argument in the sums of 
Eqs.~(\ref{eq:ev0}) and~(\ref{eq:ev1}).
Furthermore, in the Euler-Maclaurin formula, $\gamma(\omega)$ only occurs
through its derivatives of $n$-th order $\gamma^{(n)}$ at the endpoints 
$\omega=0$ and $\pi$, which satisfy $\gamma^{(n)}(\pi)=-\gamma^{(n)}(0)$,
see Eq.~(\ref{eq:gamma_r3}).
This is enough to guarantee that any coefficient $a_k(R)\ [\,b_k (R)]$ will
differ from its isotropic counterpart  $a_k(1)\ [\,b_k (1)]$ by a multiplicative
correction, $g_k(s_1)$.
Thus, it is predicted in Ref.~\onlinecite{izhu01} that
the ratios given in Eq.~(\ref{eq:akbk}) will remain unchanged. In this context,
Eqs.~(\ref{eq:c3}) and~(\ref{eq:sqrt}) reflect the (easily checkable) 
fact that  $g_1(s_1)= s_1^{-1}$, 
where for Eq.~(\ref{eq:sqrt}) one also uses $s_1\,s_2=1$ at criticality.

In order to test the robustness of the theoretical framework just expounded, 
we evaluated the third-order correction.
This is done by replacing the argument of Eqs.~(\ref{eq:ev0}) and~(\ref{eq:ev1})
by its generalized form, Eq.~(\ref{eq:gamma_r3}), and following the corresponding
effects on the $N^{-3}$ term in Eq.~(\ref{eq:e-ml}), which arise from the third-order
derivatives indicated there. One finds:   
\begin{equation}
g_3(s_1)=\frac{1}{2 s_1^2}(s_1+\frac{1}{s_1})=\frac{1}{2 s_1 t_1^2}\ ,
\label{eq:g3}
\end{equation}  
where $t_1 \equiv \tanh 2K_1$.
 
\begin{figure}
{\centering \resizebox*{3.3in}{!}{\includegraphics*{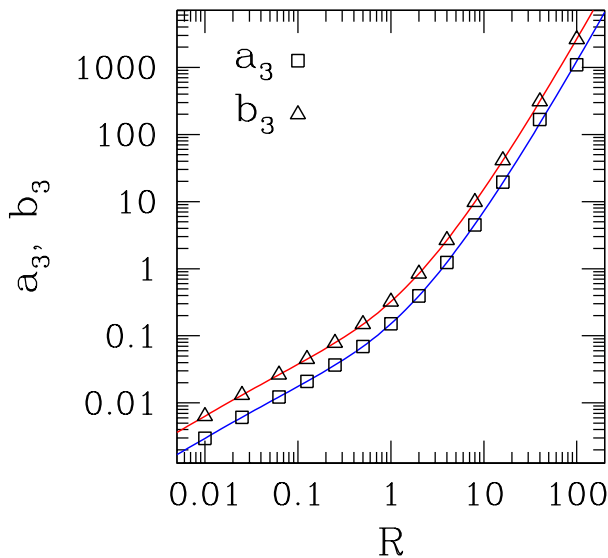}}}
\caption{(Color online) Points: results for $a_3(R)$, $b_3(R)$
defined in Eqs.~(\protect{\ref{eq:fefit}}) and~~(\protect{\ref{eq:etafit}}),
from fits of $f_N$, $\xi_N^{-1}$
data evaluated for strips of widths $100 \leq N \leq 300$ and PBC across, via 
Eqs.~(\protect{\ref{eq:fNxiN}}),~(\protect{\ref{eq:ev0}}), 
and~(\protect{\ref{eq:ev1}}). Uncertainties are smaller than symbol sizes.
Full lines show $g_3(s_1)$ defined in
Eq.~(\protect{\ref{eq:g3}}), multiplied respectively by $a_3(1)$ [blue] 
and $b_3(1)$ [red] (see text). 
} 
\label{fig:a3b3}
\end{figure}

We numerically calculated $f_N$ and $\xi_N^{-1}$ 
from Eqs.~(\ref{eq:fNxiN}),~(\ref{eq:ev0}), and~(\ref{eq:ev1})
for assorted values of $R$, and  $N=10j,\ j=2,3 \dots 30 $. The resulting sequences
were adjusted to: 
\begin{eqnarray}
f_N(R)=f_\infty(R)+\frac{1}{s_1}\frac{\pi}{12 N^2}+\frac{a_3(R)}{N^4}+
\frac{a_5(R)}{N^6}\ ,
\label{eq:fefit}\\
\xi_N^{-1}(R)= \frac{1}{s_1}\frac{\pi}{4 N}+\frac{b_3(R)}{N^3}+\frac{b_5(R)}{N^5}\ 
,\qquad\qquad\ \ 
\label{eq:etafit}
\end{eqnarray}
where $f_\infty$, $\{a_k\}$, and $\{b_k\}$ are adjustable parameters.
It is important to keep the  next-higher-order terms $a_5$ and $b_5$ 
in the truncated expansions above,
in order to improve stability for the quantities $a_3$ and $b_3$ which
are the main focus of interest here. The optimum range of $N$, large enough for
higher-order terms to have negligible influence, but not so large as to compromise
the numerical accuracy of fits (since this depends crucially 
on differences between finite-$N$ estimates of $f_N$ and $\xi_N^{-1}$)
was found to be $100 \leq N \leq 300$. In Fig.~\ref{fig:a3b3} we show 
$a_3(R)$ and $b_3(R)$, fitted via Eqs.~(\ref{eq:fefit}) and~(\ref{eq:etafit}),
for several values of $R$ spanning four orders of magnitude. The continuous
lines depict Eq.~(\ref{eq:g3}), multiplied respectively by the isotropic
values $a_3(1)=7\pi^3/1440$, $b_3(1)=\pi^3/96$~\cite{dds82,dq00,izhu01}.  
The agreement is perfect, except for $a_3(R)$ at $R \gtrsim 30$ where
reasonable convergence was only obtained upon adding the next
higher-order term, $a_7(R)/N^8$, in Eq.~(\ref{eq:fefit}). 

Our results provide direct numerical evidence that the Euler-Maclaurin scheme 
used in Ref.~\onlinecite{izhu01} is indeed applicable to Ising systems with
PBC and any finite degree of (ferromagnetic) anisotropy. Similar conclusions were 
drawn for the anisotropic Ising model with Braskamp-Kunz
boundary conditions~\cite{izyeh09}.

It is interesting to consider the above results for $R~\to~\infty$.
It is known~\cite{fs78,hb81,bg85,mh87} that the [zero-temperature] quantum Ising 
chain (QIC) in a transverse field~\cite{pfeuty} 
has a correspondence with this extreme anisotropic 
limit, via:  $f_\infty \leftrightarrow E_0$, $\xi^{-1} \leftrightarrow E_1-E_0$ etc,
where the $E_i$ are the energy levels of the quantum system. In Ref.~\onlinecite{izhu01}
the energy spectrum of the QIC with PBC 
was studied directly with help of the Euler-Maclaurin 
formula, and the corresponding ratio $b_k/a_k$ was found to obey Eq.~(\ref{eq:akbk}). 
The latter result can also be extracted from the exact expressions Eqs.~(17a) 
and~(18a) of Ref.~\onlinecite{bg85}. 

While the limit given in Eq.~(\ref{eq:sqrt}) is preserved as $R \to \infty$, 
the exponential divergence of the higher-order terms is not cancelled:  
\begin{equation}
[\,\xi_{N(1)}^{-1}\,\xi_{N(2)}^{-1}\,]^{1/2}=\frac{\pi}{4N} + \frac{1}{4}\,(s_1+
\frac{1}{s_1})^2\,\frac{\pi^3}{96 N^3} + \cdots\ .
\label{eq:sqrt2}
\end{equation} 
A similar effect (with the factor $1/s_1$) is already obvious 
in Eq.~(\ref{eq:c3}). In summary, although coefficient ratios $b_k/a_k$
are preserved as $R \to \infty$, each term of the Euler-Maclaurin
expansion for the two-dimensional Ising model with $R \gg 1$ is translated
into its counterpart of the corresponding expansion for the QIC by means of a 
distinct anisotropy factor. 
 
\section{Free boundary conditions}
\label{sec:fbc}

\subsection{Isotropic systems}
\label{isfbc}

The eigenvalue spectrum of the TM has been obtained~\cite{dba71,aks88,aks89}
for Ising $S=1/2$ strips with nearest-neighbor couplings
and  free boundary conditions (FBC) across.
For a strip of width $N$ sites, one has:
\begin{equation}
\ln \Lambda_m=\frac{1}{2}\sum_{i=1}^N \pm \gamma(\omega_i)\ ,\quad m=0, \cdots, 2^N-1,
\label{eq:evf}
\end{equation}
where the $\pm$ combinations run through all $2^N$ possibilities. A regular 
background term, $\frac{1}{2}N \ln (2 \sinh 2K_1)$ [$\,$see 
Eqs.~(\ref{eq:ev0}) and~(\ref{eq:ev1})$\,$], has been omitted. With 
all the $\gamma(\omega_i)$ real and positive for this case~\cite{dba71},
\begin{eqnarray}
\ln \Lambda_0 =\frac{1}{2}\sum_{n=1}^N  \gamma(\omega_n)\ ;
\label{eq:fe_fbc}\\
\xi_N^{-1}=\gamma(\omega_1)\ ,\qquad\quad 
\label{eq:eta_fbc}
\end{eqnarray}
where $\omega_1$ corresponds to the smallest $\gamma$.
The relationship between the $\gamma$ and the allowed frequencies $\omega_i$
is given by~\cite{dba71}:
\begin{equation}
\cosh \gamma =\cosh 2K_1^\ast\,\cosh 2K_2-\sinh 2K_1^\ast\,\sinh 2K_2\,\cos 
\omega\ ; \ \label{eq:gamma_2}\\
\end{equation}
\begin{eqnarray}
\sinh \gamma\,\cos \delta^\ast=
\sinh 2K_2\,\cosh 2K_1^\ast - \nonumber \\
-\cosh 2K_2\,\sinh 2K_1^\ast\,\cos \omega\ ;\
\label{eq:gamma_3}
\end{eqnarray}
\begin{equation}
\sin \omega / \sinh \gamma =\sin \delta^\ast / \sinh 2K_1^\ast\ .\qquad\qquad
\label{eq:gamma_4}
\end{equation}
From Eqs.~(\ref{eq:gamma_2}), (\ref{eq:gamma_3}), and ~(\ref{eq:gamma_4}),
one gets at the critical point, where $s_1\,s_2=1$:
\begin{eqnarray}
\cosh \gamma =1+\frac{1}{s_1^2}(1-\cos \omega)\ ;\qquad\qquad\
\label{eq:gamma_f}\\
\tan \delta^\ast =t_1\,\frac{\sin \omega}{(1-\cos \omega)}\ ,\qquad\quad
\label{eq:omega-delta2}
\end{eqnarray}
again with $t_1 \equiv \tanh 2K_1$.
From Eq.~(\ref{eq:gamma_f}), the smallest $\gamma$ corresponds to
the lowest allowed $\omega$. Note also that Eq.~(\ref{eq:gamma_f}) is identical in
form to Eq.~(\ref{eq:gamma_r2}), so it can also be rewritten as 
Eq.~(\ref{eq:gamma_r3}).
Finally, the allowed frequencies $\omega_n$ can be 
determined from Eq.~(\ref{eq:omega-delta2}) combined with the quantization 
condition~\cite{dba71}:
\begin{equation}
e^{i\,N\,\omega} =\pm e^{i\,\delta^\ast}\ ,\quad 0 \leq  \omega \leq \pi\ ,
\label{eq:omega-delta}
\end{equation}
by eliminating the auxiliary angle $\delta^\ast$.

For the remainder of this Subsection, we shall consider only isotropic 
systems ($K_1=K_2$), thus $s_1=1$, $t_1=1/\sqrt{2}$ in Eqs.~(\ref{eq:gamma_f})
and~(\ref{eq:omega-delta2}). 

\begin{figure}
{\centering \resizebox*{3.3in}{!}{\includegraphics*{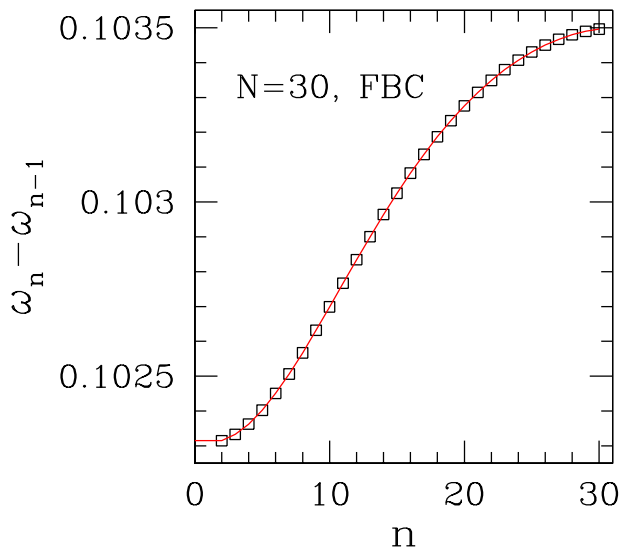}}}
\caption{(Color online)
Ising strips with FBC and isotropic interactions, at $T_c$:
differences $\omega_n-\omega_{n-1}$ ($n \geq 2$) between consecutive solutions of 
Eqs.~(\protect{\ref{eq:omega-delta2}}) and~(\protect{\ref{eq:omega-delta}}),
for strip of width $N=30$ sites.
} 
\label{fig:wdiff}
\end{figure}

The resulting frequencies are not equally spaced, as illustrated in 
Fig.~\ref{fig:wdiff}. 
So the Euler-Maclaurin formula cannot be used in the same way as in 
Ref.~\onlinecite{izhu01}, to calculate 
the free energy from Eq.~(\ref{eq:fe_fbc}). However, we show in
the following that one can
still make adaptations and extract some useful information.
We found that for large $N$ the $\omega_n$ approach the form:
\begin{equation}
\omega_n=\omega_n^0+\frac{1}{N}\,f\left(
\frac{n}{N}\right)  ,\quad \omega_n^0 \equiv 
\left(n-\frac{1}{2}\right)\frac{\pi}{N}\ ,\ 1 \leq n \leq N\ .
\label{eq:omega-fbc}
\end{equation}
\begin{figure}
{\centering \resizebox*{3.3in}{!}{\includegraphics*{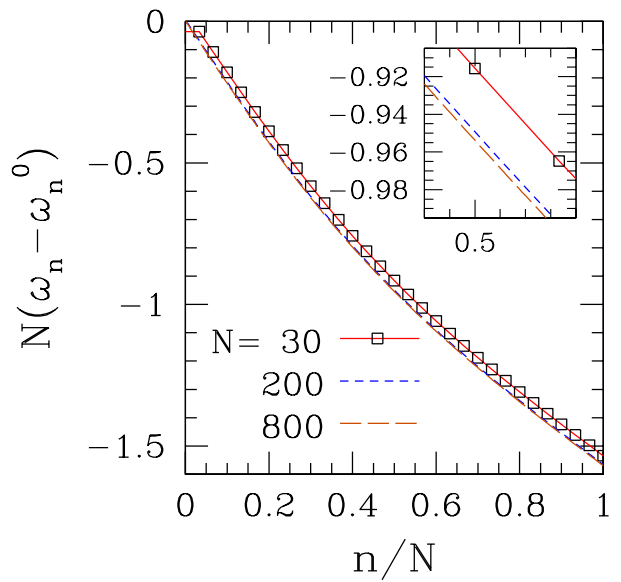}}}
\caption{(Color online)
Ising strips with FBC and isotropic interactions, at $T_c$:
illustrating the convergence of $N(\omega_n-\omega_n^0)$ towards $f(n/N)$ with
increasing $N$; see Eq.~(\protect{\ref{eq:omega-fbc}}). The inset is a blowup
of the central section of the main figure.
} 
\label{fig:scf}
\end{figure}
As shown in Fig.~\ref{fig:scf}, $f(u)$ is a smoothly varying function of $u=n/N$.
One has: $f(0)=0$, $\lim_{\,u \to 1} f(u) = -\pi/2$. Both limits can be understood
by examination of the graphical solutions of Eqs.~(\ref{eq:omega-delta2})
and~(\ref{eq:omega-delta})~\cite{aks89}.
The residual $N$-dependence of $N(\omega_n-\omega_n^0)$ is highlighted in the 
inset of Fig.~\ref{fig:scf}. This can be accounted for by an additive
correction of the form $(1/N)\,f_2(n/N)$; $f_2$ is nearly constant, 
varying
smoothly between $\approx 1.1$ and $1.3$ for $0 < n/N <1$.
We thus write:
\begin{equation}
\gamma (\omega_n)= \gamma (\omega_n^0) + 
\sum_{j=1}^\infty\frac{1}{N^j}\,g^{(j)}(\omega_n^0)\ ,
\label{eq:gamma-fbc1}
\end{equation}
with
\begin{eqnarray}
g^{(1)}(\omega_n^0) \equiv \left\{ f(\omega)\,
\gamma^{(1)} (\omega)\right\}_{\omega_n^0}\ ,\qquad\qquad
\nonumber \\
g^{(2)}(\omega_n^0) \equiv \left\{ f_2(\omega)\,\gamma^{(1)}(\omega)+
\frac{1}{2!}\,f^2(\omega)\,\gamma^{(2)}(\omega)\right\}_{\omega_n^0}\ \dots\qquad
\label{eq:gamma-fbc2}
\end{eqnarray}
where $\gamma^{(m)} \equiv d^m \gamma/d\omega^m$, and the arguments of $f$ and $f_2$
have been straightforwardly changed. Eq.~(\ref{eq:fe_fbc})
then becomes:
\begin{equation}
\ln\Lambda_0 =\frac{1}{2}\sum_{n=1}^N \gamma (\omega_n^0)+ \frac{1}{2N}\sum_{n=1}^N   
g^{(1)}(\omega_n^0) 
+\dots \equiv \sum_{j=0}^\infty S_j\ ,
\label{eq:e-ml2}
\end{equation}
So, each term (of order $j \geq 0$, with $g^{(0)}(\omega_n^0) \equiv 
\gamma(\omega_n^0)$) 
of the Taylor expansion indicated in Eq.~(\ref{eq:gamma-fbc1})
gives rise to a sum $S_j$ of $N$ terms, each of the latter
evaluated at $\omega=\omega_n^0$ ($1 \leq n \leq N$),
i.e., at equally spaced intervals.

We investigated the feasibility of applying  the Euler-Maclaurin formula,
Eq.~(\ref{eq:e-ml}), to each $S_j$, 
with $x=\omega$, $h=\pi/N$, $a=0$, $b=\pi$, $\alpha=1/2$,
so that the result would be of the form 
$S_j=\sum_{i=-1}^\infty a_i^j/N^{(i+j)}$ [$\,$where $i=-1$ corresponds to the integral
in Eq.~(\ref{eq:e-ml})$\,$]. $S_j$ would then give 
contributions to $\ln \Lambda_0$  at all orders $N^{-(j+i)}$, $i \geq -1$. 
Note that $a_0^j \equiv 0$ because $B_1(1/2)$=0~\cite{abram,ivizhu02}.
However, one would have to assume that the infinite sum implicit in each
Taylor series commutes with the infinite sum present in each separate
Euler-Maclaurin expansion (the form given in Eq.(\ref{eq:e-ml}) assumes that
the remainder term vanishes; see, e.g., Ref.~\onlinecite{ivizhu02}). Having in mind
that the expansion parameter of the Taylor series and 
the sampling interval of the Euler-Maclaurin formula can be of the same order
($\pi/N$), it is doubtful that such commutation can be guaranteed. 

With these words of caution in mind, here
we evaluate only a few of the lowest-order terms which would
occur in such a calculational framework. 

We applied the Euler-Maclaurin formula to $S_0$ in Eq.~(\ref{eq:e-ml2}).
This differs from the sum indicated in its PBC
counterpart, Eq.~(\ref{eq:ev0}), in that the frequency spacing here is 
half that in the latter Equation. For the corresponding integral of Eq.~(\ref{eq:e-ml}), 
this is compensated by the fact that the integration interval is cut in half as 
well, so  from $a_{-1}^0$ 
one reobtains the bulk result 
$f_\infty-(1/2)\ln 2=(2G/\pi)$, $G=0.915965594 \dots$ 
(Catalan's constant)~\cite{domb60}. 
For the terms of Eq.~(\ref{eq:e-ml}) involving 
derivatives of the $m$--th order, the corresponding term in $S_0$ 
has an extra factor
$2^{-(1+m)}$ relative to its PBC analogue~\cite{izhu01}. 
One gets $a_1^0=\pi/48$ as given by conformal invariance~\cite{bcn86}, 
$a_3^0=0.00942 \dots$~\cite{izhu01}.

For $S_1$, 
we evaluated $I \equiv \int_0^\pi g^{(1)}(\omega)\,d\omega$ using
finite-$N$ approximations for $f(x)$ with $200 \leq N \leq 2000$,
and extrapolating the resulting sequence against $1/N$. The final result is
$I/2\pi=a_{-1}^1=-0.1817309(1)$, 
to be compared with $2f_{\rm surf}=-0.18173148 \dots$~\cite{au-yang}. 
In the computation of higher-order terms, we ran into inconsistencies
between results thus obtained, and those coming from direct numerical evaluation
of the free energy via Eq.~(\ref{eq:fe_fbc}). We conjecture that these difficulties
stem from the conceptual problems in interchanging the order of infinite sums,
referred to above.  

As regards the correlation length, from Eq.~(\ref{eq:eta_fbc}) above, and combining 
Eqs.~(\ref{eq:eta}) and~(\ref{eq:gamma_r3}), 
one has for the finite-$N$ estimate $\eta_1^s(N)$ of the decay-of-correlations
exponent $\eta_1^s=2 x_1^s$: 
\begin{equation}
\eta_1^s(N)= \frac{2N}{\pi} \ln\, [y+\sqrt{y^2+1}]\ ,\quad y=2-\cos \omega_1\ .
\label{eq:eta_s}
\end{equation}
By solving Eqs.~(\ref{eq:omega-delta2}) and~(\ref{eq:omega-delta}) in the limit
$\omega \to 0$, $\delta^\ast \to \pi/2$, and consequently taking $y \to 1$
in Eq.~(\ref{eq:eta_s}), one gets:
\begin{equation}
\eta_1^s(N)=1 -\frac{1}{\sqrt{2}}\,\frac{1}{N}+ \left[\frac{1}{2}-
\frac{\pi^2}{48}\right]\,\frac{1}{N^2} + {\cal O}(N^{-3})\ .
\label{eq:eta_s2}
\end{equation}
According to Eq.~(\ref{eq:eta_s2}), both odd and even powers of $N^{-1}$
are predicted to arise in the expansion of $\xi_N^{-1}$ for this case.
For Ising  systems with FBC, the occurrence of $N^{-1}$ corrections to 
finite-$N$ estimates of scaling powers was noted in Ref.~\onlinecite{bg85b}.  

We evaluated $f_N$ and $\xi_N$ for $N=10j$, $j=2, \dots, 30$,
by numerically solving for the allowed frequencies and then plugging the results
into Eq.~(\ref{eq:gamma_f}) and, finally, Eq.~(\ref{eq:evf}).

We fitted free-energy data for $100 \leq N \leq 300$ to a truncated form of 
Eq.~(\ref{eq:c1}), with $k\leq 4$. After ensuring that known
quantities were reproduced to good accuracy when allowed to vary freely, 
we fixed them at their known values, 
namely $f_\infty=(1/2)\ln 2+(2G/\pi)$~\cite{domb60};
$f_{\rm surf}=-0.0908657 \dots$~\cite{au-yang}; $a_1=\pi/48$, with the results: 
$a_2=-0.04616(2)$, $a_3=0.024(1)$, $a_4=0.69(6)$. 
Note that $a_3$ as given here differs from $a_3^0$ evaluated from $S_0$
above, in connection with Eq.~(\ref{eq:e-ml2}). This is because
$a_3$ gets additional contributions from higher-order sums 
$S_k$, $k >0$ (not calculated there).

A fit of a subset ($100 \leq N \leq 300$) of the $\eta_1^s(N)$ thus 
obtained to the form $\eta_1^s(N)=\eta_1^s+\sum_{k=1}^4
b^{\,\prime}_k\,N^{-k}$ gave $\eta_1^s=1\ (\pm 1 \times 10^{-10})$, $b_1^{\,\prime}
=-0.70710680(3)$, $b_2^{\,\prime}=0.294388(7)$, $b_3^{\,\prime}=0.2274(7)$, 
$b_4^{\,\prime}=-0.68(3)$. By keeping $\eta_1^s$, $b_1^{\,\prime}$, 
$b_2^{\,\prime}$ fixed at the 
respective values predicted in Eq.~(\ref{eq:eta_s2}), we obtained
$b_3^{\,\prime}=0.227972(6)$, $b_4^{\,\prime}=-0.7013(6)$. 
The above results both confirm the predictions of Eq.~(\ref{eq:eta_s2}) for 
$b_1^{\,\prime}$ and $b_2^{\,\prime}$, and indicate that, 
in general, both even and odd powers of $N^{-1}$ 
occur in the expansion whose lowest-order  terms are given in that Equation.

We defer analysis of the ratios $b_k/a_k$ thus obtained until the next Subsection,
where anisotropic systems with FBC, and their connection to the QIC with free 
ends, are discussed.

\subsection{Anisotropic systems with FBC}
\label{sec:anisfbc}

We first note that, even though Eq.~(\ref{eq:gamma_r3}) is valid here,
the arguments given immediately below it 
do not seem to cover the present case, 
since for FBC the $\omega_n$ depend on anisotropy in the
non-trivial way given in Eqs.~(\ref{eq:omega-delta2}) 
and~~(\ref{eq:omega-delta}). Thus it is not obvious whether, e.g., 
Eq.~(\ref{eq:g3}) still applies to the free energy here.

We have directly examined the $\omega_n$, 
for varying anisotropies, and seen that their
behavior is qualitatively similar to that for the isotropic case, depicted in 
Figs.~\ref{fig:wdiff} and~\ref{fig:scf}. In particular, the limits
$f(0)=0$ and $f(1)=-\pi/2$ still hold [$\,$see the comments following 
Eq.~(\ref{eq:omega-fbc})$\,$].

By incorporating anisotropy into Eq.~(\ref{eq:eta_s}) via Eq.~(\ref{eq:gamma_r3}),
one gets the generalized version of Eq.~(\ref{eq:eta_s2}):
\begin{equation}
\eta_1^s(N)=\frac{1}{s_1}\left\{1 -\frac{1}{2 t_1}\frac{1}{N}+ \frac{1}{2 
t_1^{\,2}}\left[\frac{1}{2}-
\frac{\pi^2}{48}\right]\frac{1}{N^2}\right\} + \cdots\ .
\label{eq:eta_sfbc}
\end{equation}

\begin{figure}
{\centering \resizebox*{3.3in}{!}{\includegraphics*{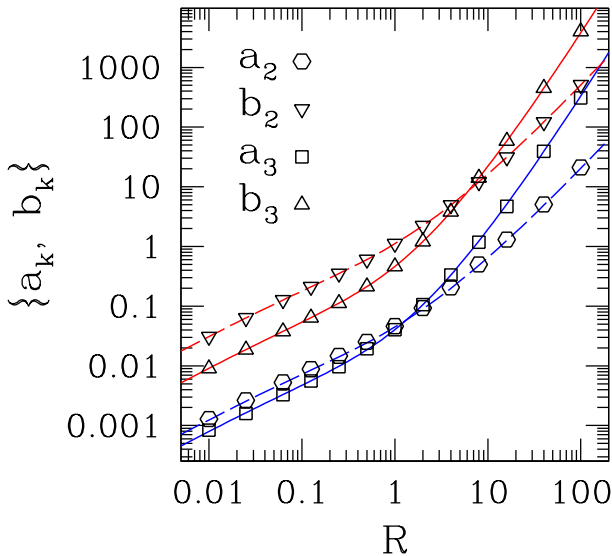}}}
\caption{(Color online) Points: results for $\{a_k(R)$, $b_k(R)\,\}$
defined in Eqs.~(\protect{\ref{eq:fefit_fbc}}) 
and~(\protect{\ref{eq:etafit_fbc}}) [$\,$absolute values for $k=2\,$], 
from fits of $f_N$, $\xi_N^{-1}$
data evaluated for strips of widths $100 \leq N \leq 300$ and FBC across.
Uncertainties are smaller than symbol sizes. Lines show anisotropy
factors  defined in Eq.~(\protect{\ref{eq:eta_sfbc}}), 
multiplied respectively by $a_k(1)$ [blue] and $b_k(1)$ [red] [$k=2,3$] (see text). 
} 
\label{fig:a2b1}
\end{figure}

We numerically calculated $f_N$ and $\xi_N^{-1}$ 
from Eqs.~(\ref{eq:fNxiN}),~(\ref{eq:fe_fbc}), and~(\ref{eq:eta_fbc})
for assorted values of $R$, and  $N=10j,\ j=2,3 \dots 30 $. 
Bearing in mind the FBC-adapted forms of 
Eqs.~(\ref{eq:c3}) and~(\ref{eq:sqrt})~\cite{inw86,nb83},
the resulting sequences were adjusted to: 
\begin{eqnarray}
N(f_N-f_\infty)=2 f_{\rm surf}+\frac{1}{s_1}\frac{\pi}{48 N}+
\sum_{k=2}^4\frac{a_k(R)}{N^k}\ ,\qquad
\label{eq:fefit_fbc}\\
\xi_N^{-1}(R)= \frac{1}{s_1}\frac{\pi}{2N}+\sum_{k=2}^4
\frac{b_k(R)}{N^k}\ ,\qquad\qquad 
\label{eq:etafit_fbc}
\end{eqnarray}
where $f_\infty$, $f_{\rm surf}$ (both of which, as well as $f_N$, also depend on $R$), 
$\{a_k\}$, and $\{b_k\}$ are adjustable parameters.
The $b^\prime_k$, defined in connection with Eqs.~(\ref{eq:eta_s})
and~(\ref{eq:eta_s2}), relate to the $b_k$ of Eq.~(\ref{eq:etafit_fbc}) by
$b_k=(\pi/2)\,b_{k-1}^\prime$.
As done in Section~\ref{sec:pbc}, we keep the  next-higher-order terms $a_4$ and $b_4$ 
in the truncated expansions above,
in order to improve stability for the quantities $a_2$, $a_3$, $b_2$,
and $b_3$ which
are the main focus of interest here. Similarly, the range of 
$N$ used in our fits was $100 \leq N \leq 300$. 
In Fig.~\ref{fig:a2b1} we show 
$a_2(R)$, $a_3(R)$, $b_2(R)$, and $b_3(R)$, 
fitted via Eqs.~(\ref{eq:fefit_fbc}) and~(\ref{eq:etafit_fbc}),
for several values of $R$ spanning four orders of magnitude. The
lines depict the anisotropy factors from Eq.~(\ref{eq:eta_sfbc}),
namely $g_2 \equiv (2s_1 t_1)^{-1}$ (dashed) and $g_3 \equiv (2s_1 t_1^2)^{-1}$ (full) 
multiplied by the pertinent values of $a_2(1)$ and $b_2(1)$
[$\,$for $g_2\,$] or $a_3(1)$ and $b_3(1)$[$\,$for $g_3\,$].    
Once again, the agreement is perfect. The only case
for which the higher-order terms ($a_4$ or $b_4$) 
made any perceptible difference  was for $a_3(R)$ at $R \gtrsim 30$. 

Our results provide direct numerical evidence that the coefficient ratios
$b_2/a_2=24.06(2)$ and $b_3/a_3=19.3(6)$ remain
constant against any finite degree of (ferromagnetic) anisotropy. 
Note that $b_3/a_3$ differs substantially from the PBC value $15/7$~\cite{izhu01}.
It is remarkable that the free-energy coefficients depend on anisotropy
in the same way as those for the correlation length. As stated in the
first paragraph of this Section, this is not obviously granted at the outset. 

We consider the extreme anisotropic limit $R \to \infty$ of Ising strips with FBC, 
and its connection to the QIC with FBC at both ends~\cite{pfeuty,bg85}. 
In Ref.~\onlinecite{izhu09}, the exact expressions for  ground-state energy and 
energy gaps of the QIC, given in Ref.~\onlinecite{pfeuty},
were written as Euler-Maclaurin expansions;
similarly to the PBC case, only odd powers of $N^{-1}$ were found to occur in 
the corresponding forms of Eqs.~(\ref{eq:c1}) and~(\ref{eq:eta1}). The counterpart
to Eq.~(\ref{eq:akbk}) in this case was shown to be:
\begin{equation}
\frac{b_k}{a_k}=\frac{2(k+1)}{(2^k-1)\,B_{k+1}}\quad (k=2j-1\ ,\ j\geq 1)\ [\,{\rm FBC
}\,]~.
\label{eq:akbk2}
\end{equation}
For $k=1$, this agrees with the conformal-invariance results
of Eqs.~(\ref{eq:c}) and~(\ref{eq:eta}); for $k=3$, Eq.~(\ref{eq:akbk2}) gives
$b_3/a_3=-240/7$~\cite{izhu09}. 

Our results above for classical Ising spins 
differ from those for the QIC in that: (i) both even and odd powers of $N^{-1}$
occur, in free-energy as well as in correlation-length expansions; and
(ii) $b_3/a_3=+19.3(6)$, incompatible
with the value given in Ref.~\onlinecite{izhu09}.

Note however that, when one considers the two-dimensional classical Ising model
with the TM running along the  diagonal~\cite{obpw96}, in the corresponding version 
of FBC, a picture closer to that found for PBC emerges, namely~\cite{izpriv}: 
only odd powers of $N^{-1}$ occur in Eqs.~(\ref{eq:c1}) and ~(\ref{eq:eta1}),
and the value $b_3/a_3=-240/7$ is reproduced.

\section{S=1}
\label{sec:s=1}

We considered $S=1$ Ising systems on a square lattice, with both PBC and FBC,
and isotropic couplings only.
The critical temperature is known rather accurately~\cite{bcg03}, 
$J/k_BT_c=K_c=0.590473(5)$.

In this case, no closed-form expressions for the TM eigenvalues are
forthcoming, so one must rely on numerical diagonalization. 
The first consequence of this fact is that the assumption of only
integer powers in Eqs.~(\ref{eq:c1}) and~(\ref{eq:eta1}) must be reanalyzed.
Indeed, while in Sections~\ref{sec:pbc} and~\ref{sec:fbc} one could
verify directly from the respective closed-form equations 
that no noninteger powers of $N^{-1}$ were allowed, here this
possibility does not arise. Furthermore, it has
been shown for models very closely related to the standard Ising model
that fractional powers occur in corrections to scaling~\cite{bf85,bdn88}.
It was conjectured that these would take the form $N^{-4/3}$, clearly
a very important term in the current context. 
However, for the $S=1$ Ising model on a square lattice, it has been 
numerically shown that the amplitude of a hypothetical $N^{-4/3}$ term is most likely 
zero~\cite{bdn88}, so in the current Section at least, 
one can retain Eqs.~(\ref{eq:c1}) and~(\ref{eq:eta1}) in their original form.    

Secondly,
the range of strip widths within practical reach is much restricted in
comparison with $S=1/2$ systems. We used $4 \leq N \leq 16$.  
Such a narrow range was, by far, the most quantitatively relevant
source of systematic inaccuracies in our estimates of corrections to scaling, 
far outweighing, e.g., the uncertainties in $T_c$.  

In order to assess the associated effects, we produced fits of free-energy
and correlation-length data for sets of $S=1/2$ data restricted to the same
range of $N$.
For PBC, we took truncated forms of Eqs.~(\ref{eq:c1}) and~(\ref{eq:eta1})
using the exact values of $f_\infty$, $a_1$, and $b_1$, with $\{a_k,b_k\}$
as adjustable parameters for $k=3,5,7$, and zero otherwise. The $k=7$ terms
were included in order to increase stability for the $k=3$ and $5$ ones.
By further restricting the range of data fitted to $10\leq N \leq 16$,
we found very good agreement with the known values~\cite{dq00,izhu01}
of  $a_3$, $b_3$, while for $a_5$ and $b_5$ deviations were of order $5\%$
(see Table~\ref{ts1}).

Turning to $S=1$ with PBC, allowing for $a_2$, $b_2 \neq 0$ in 
Eqs.~(\ref{eq:c1}) and~(\ref{eq:eta1}) gave fitted values of order $10^{-3}-10^{-4}$
(compared with $a_3$, $b_3$ of order $10^{-1}$). We take this as signalling 
that, very likely, $a_2=b_2 \equiv 0$. Taking $a_4$, $b_4 \neq 0$   
produced uncertainties of $50\%$ or more in the corresponding estimates.
This latter fact does not provide as compelling an argument to
assume $a_4=b_4=0$ as the preceding one for $a_2$, $b_2$. However,
in view of the limited number of data available for fitting, we decided that  
this was the most prudent route to take. 

Using $10\leq N \leq 16$ and proceeding as described above for $S=1/2$,
we found the results shown in the last column of Table~\ref{ts1}.
Even assuming the systematic error in this case to be two orders of
magnitude larger than that for $S=1/2$, one gets $b_3/a_3=1.50(6)$, still 
at least $10$ error bars away from encompassing the $S=1/2$ value. 
We refrain from attaching much significance to the estimates of $b_5/a_5$, due 
to the large uncertainty in $b_5$.

For $S=1$ systems with FBC, we fixed $a_1=\pi/48$~\cite{bcn86},
$b_1=\pi/2$~\cite{cardy84}. Upon extrapolation of both PBC and FBC data,
the non-universal bulk free energy is estimated as $f_\infty=1.317600(1)$.
The surface free energy is $f_{\rm surf}=-0.095187(1)$. Although the latter quantities
are immediate byproducts of TM calculations, their value for $S=1$ Ising
spins on a square lattice does not seem to be available in the published 
literature~\cite{bn85}. Free-energy fits assuming $a_2$, $a_3$, and $a_4$  as
free parameters (the latter, for the purpose of stabilization of the former two), 
$a_k \equiv 0$ for $k \geq 5$,
gave $a_2=-0.0238(2)$, $a_3=0.019(3)$, i.e., both of the same order
of magnitude, contrary to the corresponding case for PBC.
With similar assumptions for fits of correlation-length data, we obtained
$b_2=-0.5719(1)$, $b_3=-0.562(1)$. 

\begin{table}
\caption{\label{ts1}
For $S=1/2$ and $1$, coefficients $a_k$, $b_k$ from Eqs.~(\protect{\ref{eq:c1}})
and~(\protect{\ref{eq:eta1}}), and their ratios $b_k/a_k$. Columns 2 and 4: 
calculated from fits of
free-energy and correlation-length data for strips with PBC across,
and $10 \leq N \leq 16$, to truncated forms of those Equations
(see text). Quoted uncertainties refer exclusively to the fitting procedures, 
i.e., no account is taken of likely systematic errors.
Column 3: exact values from Ref.~\protect{\onlinecite{izhu01}}.
}
\vskip 0.2cm
\begin{ruledtabular}
\begin{tabular}{@{}lccc}
Type & $S=1/2$, fit & $S=1/2$, exact  & $S=1$, fit\\
$a_3$ &  $0.15082(1)$ & $0.15073 \dots$ & $0.14772(2)$ \\
$a_5$ &  $0.365(1)$ & $0.39214 \dots$ & $0.313(3)$ \\
$b_3$ &  $0.32305(1)$ & $0.322982 \dots$ & $0.2212(2)$ \\
$b_5$ &  $0.766(3)$ & $0.79692 \dots$ & $0.11(5)$ \\
$b_3/a_3$ &  $2.1420(3)$ & $2.142857 \dots$ & $1.497(2)$ \\
$b_5/a_5$ &  $2.098(3)$  & $2.032258 \dots$ & $0.35(16)$ \\
\end{tabular}
\end{ruledtabular}
\end{table}   

\section{Second-neighbor couplings}
\label{sec:nnn}

For square-lattice $S=1/2$ spins with nearest-neighbor (next-nearest neighbor) 
couplings $J$ ($J^\prime$), we considered both interactions ferromagnetic
and $J^\prime/J=1$. Again, the critical point is known to excellent 
accuracy~\cite{nb98}, $K_c=0.1901926807(2)$.

Once more, one must use numerical diagonalization of the TM since no closed-form
expressions are available for the eigenvalues. We took $4 \leq N \leq 22$, a
significantly broader range than was feasible for $S=1$ in the preceding Section,
but not in any way comparable to the leeway one has for $S=1/2$ with first-neighbor
interactions only. 

Similarly to Section~\ref{sec:s=1}, one must investigate whether noninteger
powers show up  in the corrections to scaling, $N^{-4/3}$ being a likely
candidate~\cite{bf85,bdn88}. We did this by fitting our PBC free-energy
and correlation length data respectively to:
\begin{eqnarray}
f_N=f_\infty +\frac{\pi}{12N^2}+\frac{a_{x_f}}{N^{x_f}}\ ;\nonumber \\
\xi_N^{-1}=\frac{\pi}{4N}+\frac{b_{x_\xi}}{N^{x_{\xi}}}\ ,
\label{eq:fracpow}
\end{eqnarray}
where the adjustable powers $x_f$, $x_\xi$ represent the dominant
non-universal corrections. From fits of data in the range $[N_0,22]$,
we found $x_f=3.83(2)$, $3.971(3)$, $3.981(3)$ respectively for $N_0=4$, $12$, and
$16$, and  $x_\xi=2.78(3)$, $2.921(6)$, and $2.937(3)$ for the same sequence of $N_0$.
So it is apparent that $x_f \to 4$, $x_\xi \to 3$ with increasing $N$.
Comparing with Eqs.~(\ref{eq:c1}) and ~(\ref{eq:eta1}), we conclude for the absence 
of  fractional powers such as $N^{-4/3}$ here.

For strips with PBC across, our analysis was then conducted along the lines described 
for $S=1$ in Section~\ref{sec:s=1}. Contrary to the $S=1$ case,
allowing for $a_7$, $b_7 \neq 0$ did not improve stability of lower-order
coefficients, and we decided to keep both to zero. The optimum range of widths
for our fits was now $15 \leq N \leq 22$. We found $a_3=-0.09626(6)$, 
$a_5=0.210(4)$; $b_3=-0.3305(3)$, $b_5=1.96(1)$. From this, we estimate
$b_3/a_3= 3.43(1)$, which is again at variance with the $S=1/2$ value~\cite{izhu01}
$15/7=2.14286 \dots$.

For FBC, the known universal coefficients are
$a_1=\pi/48$~\cite{bcn86}, $b_1=\pi/2$~\cite{cardy84}.
Combining PBC and FBC data, the extrapolated free energy per site
is $f_\infty =0.82926462(1)$, while the surface free energy is  
$f_{\rm surf}=-0.0895385(1)$. Estimates for these quantities are not quoted in published 
work on the next-nearest-neighbor $S=1/2$ Ising model 
using TM techniques~\cite{bn85,nb98}. We attempted 
free-energy fits, at first using $a_2$, $a_3$, and $a_4$  as free parameters, and 
$a_k \equiv 0$ for $k \geq 5$. Similarly to the PBC case, allowing $a_4$
to vary did not improve stabilization of $a_2$ or $a_3$, so we set $a_4 \equiv 0$.
We thus found $a_2=0.00994(2)$, $a_3=-0.0096(1)$.
From fits of correlation-length data, we obtained
$b_2=0.23236(3)$, $b_3=-0.529(1)$. 

\section{Discussion and Conclusions} 
\label{sec:conc}
We have examined subdominant corrections to scaling for critical Ising systems
on strip geometries. One of our main goals has been to check the extent to which
the 
constant value 
of coefficient ratios, expressed in Eq.~(\ref{eq:akbk}),
remains valid within the broader Ising universality class.

In Section~\ref{sec:pbc} we considered Ising $S=1/2$ systems, on strips with PBC 
across. We investigated the effects of anisotropic interactions, extending the 
framework introduced in Refs.~\onlinecite{inw86,nb83}, and providing 
numerical evidence
that the non-universal coefficients $a_3$ and $b_3$ of Eqs.~(\ref{eq:c1})
and~(\ref{eq:eta1}) indeed follow the prediction given by Eq.~(\ref{eq:g3}).
As a byproduct, the validity of Eq.~(\ref{eq:akbk}) has been directly verified within
four orders of magnitude of anisotropy variation for this case.

In Section~\ref{sec:fbc}, for strips of spin-$1/2$ systems with FBC along one
of the coordinate axes, we examined ways in which the non-constant frequency 
spacing in the eigenvalue spectrum can be dealt with, in order to make the
sum in Eq.~(\ref{eq:fe_fbc}) amenable to treatment via the Euler-Maclaurin
summation formula. The lowest-order terms of the resulting expansion
are shown to agree with known results. 

From the correlation-length expression, Eq.~(\ref{eq:eta_fbc}), we showed
directly that both odd and even powers of inverse strip width are expected
in corrections to scaling, and explicitly evaluated the two lowest-order
non-universal coefficients [$\,$see Eq.~(\ref{eq:eta_s2})$\,$].
Generalization to anisotropic systems is given in Eq.~(\ref{eq:eta_sfbc}),
where one can see that the first- and third order anisotropy factors
(respectively, $1/s_1$ and $1/2s_1t_1^2$) 
are the same as those for PBC [$\,$see Eqs.~(\ref{eq:c3}),~(\ref{eq:sqrt}),
and~(\ref{eq:g3})$\,$]. 

We also found numerically that the amplitude ratios $b_k/a_k$ remain 
constant, 
for $k=2$ and $3$, upon introduction of anisotropic couplings. 

Sections~\ref{sec:s=1} and~\ref{sec:nnn} deal respectively with $S=1$
systems with first-neighbor interactions, and spin-$1/2$ ones with both first- and
second-neighbor couplings. For PBC we find that, in both cases, the ratio $b_3/a_3$
differs considerably from the value $15/7=2.142857 \dots$ found in
Ref.~\onlinecite{izhu01} for $S=1/2$, first-neighbor couplings only.
We quote $b_3/a_3=1.50(6)$ for the former, and $3.43(1)$ for the latter.
For FBC, comparison of $b_k/a_k$ ratios with those pertaining to $S=1/2$ systems
gives $b_2/a_2=24.0(2)$ ($S=1$), $23.4(1)$ (next-nearest-neighbor), $24.06(2)$
($S=1/2$, first-neighbor). Although the error bars do not quite overlap,
it appears that 
a constant value
of this ratio cannot be definitely discarded.
However, no such regularity is seen for $b_3/a_3$, its value being respectively 
$-30(5)$, $+55(1)$, and $+19.3(6)$ in each case.

Overall, it seems that both even and odd powers of $N^{-1}$ always show up
in Eqs.~(\ref{eq:c1}) and ~(\ref{eq:eta1}), 
for critical Ising strips with FBC along one coordinate axis. On the other hand,
for PBC only odd ones occur. Concurring remarks can be found in the
literature~\cite{dds82,bg85b}; however, it seems difficult to prove
such a statement rigorously. So far, one has to rely on case-by-case
analyses, as was done here. As pointed out at the end of Section~\ref{sec:fbc},
considering the version of FBC with the TM running along the 
diagonal~\cite{obpw96} is enough to restore a picture very similar to
that holding for PBC~\cite{izpriv}. Thus, the behavior of subdominant
corrections to scaling is sensitive to what might appear to be a minor
technical detail.

The constant value
of amplitude ratios is maintained
upon varying anisotropy for $S=1/2$ systems with first-neighbor couplings,
either with PBC or FBC; however, it does not seem to survive
changes in spin $S$, or introduction of further neighbor interactions.
We have thus established that the observed, apparently universal,
constant  amplitude ratios
pertain to a limited subset of systems which are in the broader
Ising universality class.
It remains to be further investigated whether the close values found for
 $b_2/a_2$  with FBC in the three cases are indeed an indication of
an actual constant ratio.

\begin{acknowledgments}
The author thanks N.Sh. Izmailian for extensive and illuminating correspondence,
and for pointing out Refs.~\onlinecite{obpw96,ivizhu02};
D. B. Abraham and J. L. Cardy for helpful discussions; and Helen Au-Yang
for clarification regarding Ref.~\onlinecite{au-yang}. Thanks are due also to 
the Rudolf Peierls Centre for Theoretical Physics, Oxford, for the hospitality, 
and CAPES for funding the author's visit. 
The research of S.L.A.d.Q. is financed 
by the Brazilian agencies CAPES (Grant No. 0940-10-0),  
CNPq  (Grant No. 302924/2009-4), and FAPERJ (Grant No. E-26/101.572/2010).
\end{acknowledgments}

\end{document}